\documentclass[twocolumn]{revtex4}%
\usepackage{amsmath}
\usepackage{graphicx}
\usepackage{amsfonts}
\usepackage{amssymb}%
\begin{document}
\title{Stokes-Einstein diffusion of colloids in nematics}
\author{Fr\'{e}d\'{e}ric Mondiot, Jean-Christophe Loudet, Olivier Mondain-Monval,
Patrick Snabre}
\affiliation{Centre de Recherche Paul Pascal, Universit\'{e} de Bordeaux \& CNRS, 33600
Pessac, France}
\author{Alexandre Vilquin, Alois W\"{u}rger}
\affiliation{Laboratoire Ondes et Mati\`{e}re d'Aquitaine, Universit\'{e} de Bordeaux \&
CNRS, 33405 Talence, France}

\begin{abstract}
We report the first experimental observation of anisotropic diffusion of
polystyrene particles immersed in a lyotropic liquid crystal with two
different anchoring conditions. Diffusion is shown to obey the Stokes-Einstein
law for particle diameters ranging from 190 nm up to 2 $\mu$m. In the case of
prolate micelles, the beads diffuse four times faster along the director than
in perpendicular directions, $D_{\Vert}/D_{\bot}\approx4$. In the theory part
we present a perturbative approach to the Leslie-Ericksen equations and relate
the diffusion coefficients to the Miesovicz viscosity parameters $\eta_{i}$.
We provide explicit formulae for the cases of uniform director field and
planar anchoring conditions which are then discussed in view of the data. As a
general rule, we find that the inequalities $\eta_{b}<\eta_{a}<\eta_{c}$,
satisfied by various liquid crystals of rodlike molecules, imply $D_{\Vert
}>D_{\bot}$.


\end{abstract}
\maketitle

Dispersions of colloids in nematic liquid crystals (NLCs) show singular
properties, that are related to the anisotropy of the nematic phase and to the
anchoring of the nematogens on the particle surface
\cite{Ruh96,Pou97,Sta01,Bai11}. The colloid imposes on the neighboring LC
molecules an orientation that locally breaks the uniform nematic alignment and
gives rise to elastic interactions. In order to satisfy the global boundary
conditions, each inclusion is accompanied by topological defects that
determine the long-range deformation field and govern colloidal pattern
formation \cite{Ruh97,Pou98,Lub98,Sta99,Lou00,Mus06,Koe09}.

A Brownian particle in a NLC thus drags a nematic deformation along its random
trajectory, and its diffusion behavior constitutes a sensitive probe to the
local order parameter and surface anchoring. In an isotropic medium, the
Stokes-Einstein coefficient $D=k_{B}T/3\pi\eta d$ is given by the particle
diameter $d$ and the scalar viscosity $\eta$. A more complex situation arises
in a nematic, where viscosity is a tensor quantity and where the director and
velocity fields exert forces on each other \cite{deG93,Osw05}. Thus the
viscous stress of a diffusing particle on the surrounding fluid, is not the
same for motion parallel and perpendicular to the director, resulting in two
coefficients $D_{\Vert}$ and $D_{\bot}$.\ 

So far anisotropic colloidal diffusion has been studied in thermotropic NLCs,
made of rod-like organic molecules the anchoring of which is determined by the
surface chemistry \cite{Lou04,Sma06,Tak08,Koe09a,Ska10}. Rather generally,
diffusion turns out to be faster along the director, with a ratio $D_{\Vert
}/D_{\bot}$ smaller than 2.\ A recent study on silica beads dispersed in
nematic 5CB with normal anchoring conditions \cite{Ska10}, reported that the
diffusion coefficients $D_{\Vert}$ and $D_{\bot}$ of large particles show the
size dependence $D\propto1/d$ expected from the Stokes-Einstein relation, yet
surprisingly saturate for smaller particles at an effective hydrodynamic
diameter of about $300$ nm. The role of surface chemistry, which controls the
anchoring of the nematogens, was put forward in an attempt to rationalize the
observations.
%TCIMACRO{\FRAME{ftbpFU}{8.5624cm}{6.401cm}{0pt}{\Qcb{(a) Brownian trajectory
%of a 190nm-diameter fluorescent PS particle, consisting of 2800 time steps of
%0.3 s. Diffusion is faster parallel to the director (double arrow). Inset:
%Schematic of the director field distortions around a sphere in planar
%anchoring conditions. The black dots symbolize \textquotedblleft
%boojum\textquotedblright\ defects \cite{Sta01}. (b) Histogram of the measured
%particle displacement $\delta$ parallel and perpendicular to the director
%during a time $\tau=0.3$ s, as obtained from a sample of 20,000 trajectory
%steps. The solid lines are Gaussian fits. }}{}{fig1.wmf}%
%{\special{ language "Scientific Word";  type "GRAPHIC";
%maintain-aspect-ratio TRUE;  display "USEDEF";  valid_file "F";
%width 8.5624cm;  height 6.401cm;  depth 0pt;  original-width 2.2667in;
%original-height 1.7685in;  cropleft "0.0702";  croptop "0.9100";
%cropright "0.9297";  cropbottom "0.0899";
%filename 'Fig1.wmf';file-properties "XNPEUR";}}}%
%BeginExpansion

%%%%%%%%%%%%%%%%%%%%%%%%%%%%%%%%%%%%%%%%%%%%%%%%%%%%
\begin{figure}[ptb]
\includegraphics[width=\columnwidth]{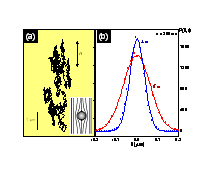}
\caption{(a) Brownian trajectory of a 190nm-diameter fluorescent PS particle,
consisting of 2800 time steps of 0.3 s. Diffusion is faster parallel to the
director (double arrow). Inset: Schematic of the director field distortions
around a sphere in planar anchoring conditions. The black dots symbolize
\textquotedblleft boojum\textquotedblright\ defects \cite{Sta01}. (b)
Histogram of the measured particle displacement $\delta$ parallel and
perpendicular to the director during a time $\tau=0.3$ s, as obtained from a
sample of 20,000 trajectory steps. The solid lines are Gaussian fits. }
\end{figure}
%%%%%%%%%%%%%%%%%%%%%%%%%%%%%%%%%%%%%%%%%%%%%%%%%%%%%

Since boundary conditions are of paramount importance in nematic colloids, we
found it worthy to investigate particle mobility in lyotropic liquid crystals
for which various anchoring conditions are easily achieved without altering
the surface chemistry \cite{Pou99}. The latter does indeed influence the
particle diffusion coefficients as shown in Ref. \cite{Koe09a}. Lyotropic LCs
are water-based surfactant mixtures, and in such systems, anchoring conditions
depend critically on the shape of the surfactant micelles (nematogens). The
latter can be tuned, through tiny changes of surfactant concentrations, from
rodlike (Calamitic Nematics, NC phase) to disklike (Discotic Nematics, ND
phase) \cite{Ama87,Qui92}. And for entropic reasons \cite{Pon88}, spontaneous
planar (NC) and normal (ND) anchoring conditions can be achieved at constant
surface chemistry (and consequently constant anchoring strength), in contrast
with the thermotropic case where varying the anchoring conditions requires a
change in the particles surface chemistry.

In this Letter, we report on both experimental and theoretical work on
anisotropic diffusion. We present data for particles with diameters $d$
ranging from 190nm to 1.9$\mu$m, both in planar and normal anchoring
conditions of the lyotropic LC at the particle surface. To the best of our
knowledge, these are the first data obtained in lyotropic systems and in both
anchoring conditions. In the theory part, we develop an original perturbative
approach which provides the effective viscosities to linear order in terms of
the Leslie coefficients. The obtained expressions enable a direct comparison
with the experimental relevant quantities and can account for the observed
diffusion anisotropy.

The two NLC phases used are the NC and the ND phases of the
water/decanol/sodium dodecyl sulfate lyotropic system, which can be obtained
at very close experimental concentrations (NC 71/24.5/4.5 \%; ND 73/23.5/3.5
\%) \cite{Ama87,Qui92}. In the NC phase, the surfactant molecules form
nanometer-sized rodlike (prolate) micelles (with long and short axes of 9 nm
and 3.5 nm \cite{Nes10}), which, in order to minimize their excluded volume,
show planar anchoring at the surface of the dispersed polystyrene spheres
(PS). In the ND phase, and for similar reasons, the disklike (oblate) micelles
(with diameter 8 nm and thickness 3.5 nm \cite{Nes10}) anchor normally at the
surface (with the disk normal perpendicular to the surface) \cite{Pou99}.
Unlike colloidal suspensions in thermotropic NLCs, this dispersion does not
require surface functionalization.%

%TCIMACRO{\FRAME{ftbpFU}{9.0215cm}{5.9353cm}{0pt}{\Qcb{Stokes-Einstein
%evolution of the diffusion coefficients $D_{\Vert}$ (along the nematic
%director) and $D_{\bot}$ (perpendicular to the nematic director) as a function
%of the particle diameter. The ratio $D_{\Vert}/D_{\bot}$ is about equal to 4.
%}}{}{fig2.wmf}{\special{ language "Scientific Word";  type "GRAPHIC";
%maintain-aspect-ratio TRUE;  display "USEDEF";  valid_file "F";
%width 9.0215cm;  height 5.9353cm;  depth 0pt;  original-width 2.6135in;
%original-height 1.8438in;  cropleft "0.0750";  croptop "0.8935";
%cropright "0.9249";  cropbottom "0.1064";
%filename 'Fig2.wmf';file-properties "XNPEUR";}}}%
%BeginExpansion

%%%%%%%%%%%%%%%%%%%%%%%%%%%%%%%%%%%%%%%%%%%%%%%%%%%%
\begin{figure}[ptb]
\includegraphics[width=\columnwidth]{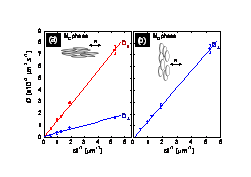}
\caption{Stokes-Einstein evolution of the diffusion coefficients $D_{\Vert}$
(along the nematic director) and $D_{\bot}$ (perpendicular to the nematic
director) as a function of the particle diameter. The ratio $D_{\Vert}%
/D_{\bot}$ is about equal to 4. }
\end{figure}
%%%%%%%%%%%%%%%%%%%%%%%%%%%%%%%%%%%%%%%%%%%%%%%%%%%%%

We used classical and fluorescence optical microscopy combined with standard
video tracking routines \cite{Cro96} to probe the Brownian diffusion in
lyotropic phases. A typical trajectory of a 190nm-diameter particle derived
from 2,800 snapshots is shown in Fig. (1a) for the NC case. Its elongation
along the nematic director $\mathbf{n}$ indicates that diffusion is faster in
this direction. In the histogram of Fig. (1c), we plot the displacements
parallel and perpendicular to $\mathbf{n}$, for a total of 20,000 trajectory
steps. The probability that the particle moves a distance $\delta$ in time
$\tau$, $P(\delta,\tau)$, is very well fitted by a Gaussian distribution; its
standard deviation is related to the diffusion coefficient through
$\overline{\delta^{2}}-\overline{\delta}^{2}=2D\tau$ \cite{Cha95}. In Fig.
(2a), we plot $D_{\Vert}$ and $D_{\bot}$ as a function of the inverse particle
radius in the NC phase whereas Fig. (2b) displays the results for the ND
phase. However, due to experimental limitations \cite{noteND}, only $D_{\bot}$
could be determined in the latter case. The straight lines confirm the linear
dependence of the Stokes-Einstein relation whatever the anchoring conditions.
The friction coefficients are different for motion parallel and perpendicular
to $\mathbf{n}$ and the large anisotropy ratio $D_{\Vert}/D_{\bot}\approx4$
indicates a strong nemato-hydrodynamic coupling in the LC matrix. Note also
the very close values measured for $D_{\bot}$ in the ND phase and $D_{\Vert}$
in the NC phase. Our results then differ considerably from the measurements of
\cite{Ska10} in a themotropic NLC, where the diffusion coefficients become
constant for particles smaller than about 300 nm, with $D_{\Vert}/D_{\bot
}\approx1.6$.

In the remainder of this paper, we study how the diffusion anisotropy arises
from the viscous properties of a NLC. The fluctuation-dissipation theorem
relates the friction coefficient to the velocity $u=F/3\pi\eta d$ of a
spherical particle driven by an external force $F$. Thus calculating the
Rayleigh function $\Psi=Fu$ provides an effective viscosity in the form
$\Psi=3\pi\eta du^{2}$, which takes two values $\eta_{\parallel}$ and
$\eta_{\perp}$ for motion parallel and perpendicular to the director.

The friction coefficients are calculated from the Leslie-Ericksen equations of
nemato-hydrodynamics for $|\mathbf{n}|=1$ and an incompressible fluid. Energy
dissipation occurs through two channels \cite{deG93,Osw05},
\begin{equation}
\Psi=\int dV\psi,\ \ \ \ \psi=\mathbf{\sigma}:\mathbf{A}+\mathbf{h}%
\cdot\mathbf{N}, \label{4}%
\end{equation}
where $\mathbf{A}$ and $\mathbf{N}$ are thermodynamic fluxes, and
$\mathbf{\sigma}$ and $\mathbf{h}$ the corresponding forces. The rate of
strain tensor%
\begin{equation}
A_{ij}=\frac{1}{2}(\partial_{i}v_{j}+\partial_{j}v_{i}) \label{6}%
\end{equation}
is given by the symmetrized derivatives of the flow $\mathbf{v}(\mathbf{r})$
in the vicinity of the particle moving at velocity $\mathbf{u}$. The vector
quantity
\begin{equation}
\mathbf{N}=\left(  (\mathbf{v}-\mathbf{u})\cdot\mathbf{\nabla}\right)
\mathbf{n-\omega\times n,} \label{8}%
\end{equation}
with the curl $\mathbf{\omega=}\frac{1}{2}\mathbf{\nabla\times v}$, expresses
the rate of change of the director with respect to the background fluid. The
conjugate forces, that is the viscous stress tensor $\mathbf{\sigma}$ and the
molecular field $\mathbf{h}$, are linear functions of the components of
$\mathbf{A}$ and $\mathbf{N}$ \cite{deG93,Osw05}. Inserting their steady-state
expressions in (\ref{4}) one has
\begin{align}
\psi &  =\alpha_{1}(\mathbf{n}\cdot\mathbf{A}\cdot\mathbf{n})^{2}+\left(
\alpha_{3}-\alpha_{2}\right)  \mathbf{N}^{2}\nonumber\\
&  \ \ \ \ \ \ +\left(  \alpha_{3}+\alpha_{2}+\alpha_{6}-\alpha_{5}\right)
\mathbf{n}\cdot\mathbf{A}\cdot\mathbf{N}\nonumber\\
&  \ \ \ \ \ \ +\alpha_{4}\mathbf{A}:\mathbf{A}+(\alpha_{5}+\alpha
_{6})\mathbf{n}\cdot\mathbf{A}\cdot\mathbf{A}\cdot\mathbf{n.} \label{14}%
\end{align}
Parodi's relation $\alpha_{3}+\alpha_{2}=\alpha_{6}-\alpha_{5}$ reduces the
viscosity tensor to five independent Leslie coefficients $\alpha_{i}$. The
various scalar products result in an intricate dependence on the relative
orientation of the macroscopic director $\mathbf{n}_{0}$ and the particle
velocity $\mathbf{u}$.

In the absence of nematic ordering, $\mathbf{n}=0$, the power density reduces
to $\psi=\alpha_{4}\sum_{ij}A_{ij}^{2}$. The tensor (\ref{6}) is readily
calculated from the velocity field of a spherical particle moving in an
isotropic liquid,%
\begin{equation}
\mathbf{v}=\left(  \frac{3a}{4r}\left(  1+\mathbf{\hat{r}\hat{r}}\right)
+\frac{a^{3}}{4r^{3}}\left(  1-3\mathbf{\hat{r}\hat{r}}\right)  \right)
\cdot\mathbf{u}\, \label{13}%
\end{equation}
where $\mathbf{\hat{r}}=\mathbf{r}/r$. The resulting Rayleigh function
$\Psi_{0}=\frac{3}{2}\pi d\alpha_{4}u^{2}$ defines the isotropic viscosity
$\eta_{0}=\frac{1}{2}\alpha_{4}$. In a NLC, however, the velocity and director
fields depend on each other through the equations for $\mathbf{\sigma}$ and
$\mathbf{h}$. Then $\Psi$ is a complicated function of the Leslie
coefficients, and it is not possible to single out the dissipation due to a
given term. Though the problem can be solved with considerable numerical
effort \cite{Ruh96,Sta01,Fuk04,Car08,Zho08,Mor11}, the resulting numbers for
the effective viscosities give no physical insight in the underlying mechanism.

\begin{table}[ptb]
\caption{Effective viscosities for zero anchoring (uniform director). The
middle column is obtained from Eqs. (\ref{18}) and (\ref{20}) with the Leslie
coefficients of 5CB and MBBA \cite{deG93,Osw05}, the last one is derived from
Stark's numerical calculations \cite{Sta01}. }%
\label{Table2}
\begin{tabular}
[c]{|l|c|c|c|}\hline
\multicolumn{2}{|l|}{Uniform director} & present work & numerically
exact\\\hline
5CB & $\eta_{\parallel}$ (P) & $0.429$ & $0.381$\\\hline
& $\eta_{\perp}$ (P) & $0.724$ & $0.754$\\\hline
MBBA & $\eta_{\parallel}$ (P) & $0.412$ & $0.380$\\\hline
& $\eta_{\perp}$ (P) & $0.650$ & $0.684$\\\hline
\end{tabular}
\end{table}

The present work\ relies on two approximations. First, we evaluate (\ref{6})
and (\ref{8}) with the above velocity field $\mathbf{v}(\mathbf{r})$ of a
particle in an isotropic liquid. Formally this amounts to linearize $\psi$
with respect to the $a_{i}$. Second we use a simple parameterization for the
director $\mathbf{n}(\mathbf{r})$ which is independent of the velocity field
and which depends on the particle size through the reduced distance $d/r\,$
only; in other words, the director has no intrinsic length scale. With these
assumptions, the Rayleigh function becomes linear in the $\alpha_{i}$ and in
the particle size $d$.

\textit{Uniform director} (UD).\ We start with the case where the particle
surface does not affect the liquid crystal order parameter. Then the director
is constant, $\mathbf{n=n}_{0}$, and with the explicit form of the tensor
$\partial_{i}v_{j}$ \cite{Lan87}, the dissipation function can be calculated
in closed form. It turns out convenient to rewrite the Leslie parameters
$\alpha_{2},...,\alpha_{6}$ in terms of the Miesovicz viscosities $\eta
_{a},\eta_{b},\eta_{c}$ given in Fig. 3; the effective viscosity for a
particle moving along the nematic order, $\mathbf{u\parallel n}_{0}$, then
reads
\begin{equation}
\eta_{\parallel}^{UD}=\frac{8\alpha_{1}}{70}+\frac{4\eta_{b}+\eta_{c}}{5}\,.
\label{18}%
\end{equation}
Similarly, we find for the perpendicular case $\mathbf{u\bot n}_{0}$
\begin{equation}
\eta_{\perp}^{UD}=\frac{3\alpha_{1}}{70}+\frac{5\eta_{a}+\eta_{b}+4\eta_{c}%
}{10}\,. \label{20}%
\end{equation}
It is noteworthy that the five independent viscosities of (\ref{14}) reduce to
three or four terms. In Table I we\ compare our formulae with Stark's
numerical calculations for the liquid crystals 5CB and MBBA \cite{Sta01}, and
find that the numbers differ by hardly 10\%. Though it slightly underestimates
the viscosity anisotropy, the linearization approximation is therefore
quantitatively correct.%

%TCIMACRO{\FRAME{ftbpFU}{3.2932in}{1.2938in}{0pt}{\Qcb{Schematic view of a
%colloidal particle in a liquid crystal. The particle moves along the $z$-axis;
%the shear in the plane $z=0$ is indicated by the decay of the fluid velocity
%field. In the left panel the director is parallel to the particle velocity,
%with an effective shear viscosity $\eta_{b}$.\ The remainder shows the
%perpendicular case $\QTR{bf}{n}_{0}=\QTR{bf}{e}_{x}$; in the middle a view of
%the $x$-$z$-plane with $\eta_{c}$, at right the $y$-$z$-plane with $\eta_{a}$.
%The Miesovicz viscosities $\eta_{a},\eta_{b},\eta_{c}$ are expressed in terms
%of the Leslie parameters. }}{}{fig3.wmf}%
%{\special{ language "Scientific Word";  type "GRAPHIC";
%maintain-aspect-ratio TRUE;  display "USEDEF";  valid_file "F";
%width 3.2932in;  height 1.2938in;  depth 0pt;  original-width 8.5314in;
%original-height 3.333in;  cropleft "0";  croptop "1";  cropright "1";
%cropbottom "0";  filename 'Fig3.wmf';file-properties "XNPEU";}} }%
%BeginExpansion

%%%%%%%%%%%%%%%%%%%%%%%%%%%%%%%%%%%%%%%%%%%%%%%%%%%%
\begin{figure}[ptb]
\includegraphics[width=\columnwidth]{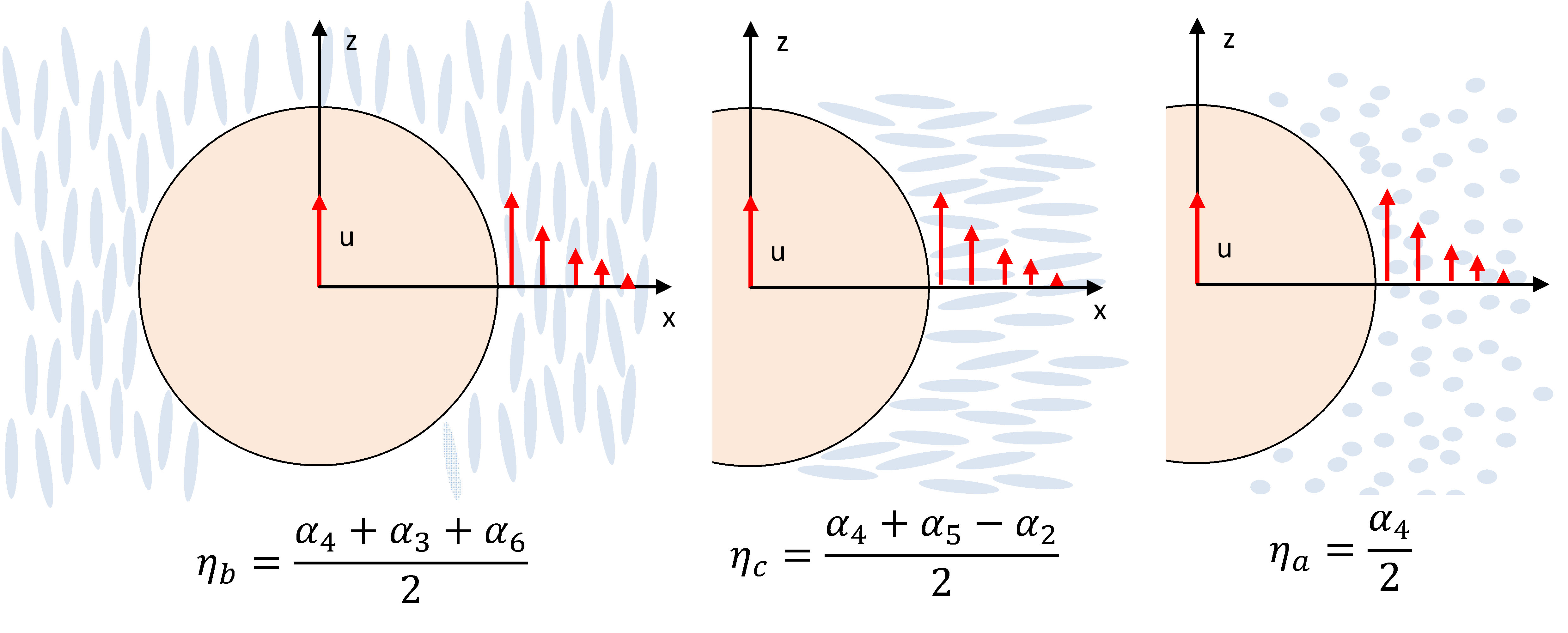}
\caption{Schematic view of a colloidal particle in a liquid crystal. The
particle moves along the $z$-axis; the shear in the plane $z=0$ is indicated
by the decay of the fluid velocity field. In the left panel the director is
parallel to the particle velocity, with an effective shear viscosity $\eta
_{b}$.\ The remainder shows the perpendicular case $\mathbf{n}_{0}%
=\mathbf{e}_{x}$; in the middle a view of the $x$-$z$-plane with $\eta_{c}$,
at right the $y$-$z$-plane with $\eta_{a}$. The Miesovicz viscosities
$\eta_{a},\eta_{b},\eta_{c}$ are expressed in terms of the Leslie parameters. }
\end{figure}
%%%%%%%%%%%%%%%%%%%%%%%%%%%%%%%%%%%%%%%%%%%%%%%%%%%%%

Because of the small weight of the first term in (\ref{18}) and (\ref{20}),
the anisotropy arises mainly from the Miesovicz viscosities. Its physical
origin is illustrated in Fig. 3 for a particle moving in the vertical
direction.\ The pole regions being of minor importance, we focus on the shear
flow in the plane $z=0$, where the viscous stress simplifies to the in-plane
derivatives of the vertical velocity component $\partial v_{z}/\partial x$ and
$\partial v_{z}/\partial y$. The left panel shows the parallel case, and the
middle and right panels the perpendicular one, with the corresponding shear
viscosities. This qualitative picture is confirmed by the large weight of
$\eta_{b}$ in the parallel viscosity (\ref{18}), and of $\eta_{a}$ and
$\eta_{c}$ in (\ref{20}). Data for common NLCs made of rodlike molecules
suggest that $\alpha_{1}$ is small; more importantly, they satisfy the
inequalities $\eta_{b}<\eta_{a}<\eta_{c}$ \cite{Osw05} and thus imply
$\eta_{\parallel}<\eta_{\perp}$, which is in line with our results.

\textit{Planar anchoring} (PA). A finite surface energy deforms the nematic
order parameter in the vicinity of the colloidal particle. Here we consider
the case of planar anchoring, which is illustrated in the inset of Fig. 1a. As
the distance from the particle increases, $\mathbf{n}$ varies smoothly toward
the constant $\mathbf{n}_{0}$. Even for the simplified one-constant elastic
energy, there is no general solution for the spatially varying order parameter
\cite{Sta01}. It is conveniently parameterized by
\begin{equation}
\mathbf{n=\mathbf{n}}_{0}\cos\Theta-\mathbf{n}_{\perp}\sin\Theta\,, \label{17}%
\end{equation}
where $\mathbf{n}_{\perp}$ is a radial vector perpendicular to $\mathbf{n}%
_{0}\,$. Because of its rotational symmetry, the director is determined by a
single function\ $\Theta(\mathbf{r})$, which decays as $1/r^{3}$ at large
distances. Here we use the ansatz of Lubensky et al. \cite{Lub98}, which for
planar anchoring results in $\Theta=\sum_{k=1}^{\infty}[\sin(2k\theta
_{n})/k](d/2r)^{1+2k}$, where $\theta_{n}$ is the polar angle with respect to
$\mathbf{n}_{0}$, that is $\cos\theta_{n}=\mathbf{\hat{r}}\cdot\mathbf{n}%
_{0}\,$.

With the director (\ref{17}) and the velocity (\ref{13}), the dissipation
function (\ref{4}) is calculated numerically for both parallel and
perpendicular alignement; it yields to the following viscosities
\begin{equation}
\eta=C_{1}\alpha_{1}+(1-C_{b}-C_{c})\eta_{a}+C_{b}\eta_{b}+C_{c}\eta_{c}%
+C_{d}\eta_{d}\,, \label{16}%
\end{equation}
where we have defined a fifth independent parameter $\eta_{d}=\frac{1}%
{2}(\alpha_{6}-\alpha_{3})=\frac{1}{2}(\alpha_{5}+\alpha_{2})\,$. From the
numerical coefficients in Table II it is clear that the viscosity anisotropy
arises essentially from $C_{b}$ and $C_{d}$. Since $C_{d}=0$ for a uniform
director, finite values of $C_{d}$ reflect distortions due to anchoring. In
the following, we discard the first term because of the small coefficient
$C_{1}$. (Moreover, $\alpha_{1}$ is also small for several thermotropic NLCs
\cite{deG93,Osw05}).

\begin{table}[ptb]
\caption{Coefficients of the effective viscosity (\ref{16}) for a spherical
particle with planar anchoring (PA), moving parallel or perpendicular to the
director. The measured values are obtained by fitting the straight lines in
Fig. (2a) with $D=k_{B}T/3\pi\eta d\,$. }%
\begin{tabular}
[c]{|c||c|c|c|c||c|}\hline
PA & $C_{1}$ & $C_{b}$ & $C_{c}$ & $C_{d}$ & measured\\\hline\hline
$\eta_{\parallel_{{}}}$ & $0.08$ & \multicolumn{1}{||c|}{$1.02$} & $0.40$ &
$-0.48$ & 0.31 Pa.s\\\hline
$\eta_{\bot_{\ }}^{^{\ }}$ & $0.04$ & \multicolumn{1}{||c|}{$0.24$} & $0.41$ &
$-0.15$ & 1.24 Pa.s\\\hline
\end{tabular}
\end{table}

\textit{Comparison with experiment}. From the straight lines in Fig. (2a) (NC
phase, planar anchoring) and the Stokes-Einstein relation, we deduce the
experimental values $\eta_{\Vert}=0.31$ Pa.s and $\eta_{\bot}=1.24$ Pa.s.
Using a plate-cone rheometer, the zero shear effective viscosity was found to
be $\eta_{S}\simeq0.4$ Pa.s at $T=25^{\circ}$C. Though these numbers are not
sufficient to determine all Leslie parameters, they present several noteworthy
constraints. In view of the coefficients listed in Table II and Eq.
(\ref{16}), we deduce $\Delta\eta=\eta_{\parallel}-\eta_{\perp}\simeq
(C_{b}^{\parallel}-C_{b}^{\perp})(\eta_{b}-\eta_{a})+(C_{d}^{\parallel}%
-C_{d}^{\perp})\eta_{d}\,$. From experiments, $\Delta\eta<0$, which therefore
implies $\eta_{b}<\eta_{a}$ and suggests that $\eta_{d}$ is small or positive.
The zero shear effective viscosity is given by $\eta_{S}=m_{a}\eta_{a}%
+m_{b}\eta_{b}+m_{c}\eta_{c}$, with $m_{i}=\frac{1}{3}$ in a polycrystalline
sample. Because of the planar anchoring conditions on the confining surfaces
of the rheometer, we expect a smaller weight for $\eta_{c}\,$; in addition,
shear-induced alignment would reduce $m_{a}$. Indeed, we find that the three
equations for $\eta_{\Vert}$, $\eta_{\bot}$, and $\eta_{S}$ have solutions
only if $m_{a,c}<0.15$ and $m_{b}>0.7$, and strongly suggest $\eta_{a}%
<\eta_{c}$. These inequalities are satisfied by the Miesovicz parameters that
are required to fit our data: For example, setting $\eta_{d}=0$ and
$m_{a,c}=0.05\,$, the measured viscosities are met with $\eta_{b}=0.27$ Pa.s,
$\eta_{a}=1.46$ Pa.s, $\eta_{c}=1.61$ Pa.s. This discussion qualitatively
agrees with that of $\eta_{\bot}$ measured for planar anchoring in the
thermotropic NLC 5CB \cite{Koe09a}.

Finally, we close with a remark on diffusion in discotic nematics (ND phase)
where the anchoring is normal and not planar anymore. As aforesaid, the data
for the perpendicular coefficient $D_{\perp}$ in Fig. (2b) are almost
identical to those for $D_{\parallel}$ in the NC phase. As discussed above,
the left panel of Fig. 3 implies $\eta_{\parallel}\sim\eta_{b}$ for prolate
micelles; a similar argument for oblate ones suggests $\eta_{\perp}\sim
\eta_{c}\,$. Consequently, the important point is that, in the ND phase, one
expects the Miesovicz viscosities to satisfy $\eta_{c}<\eta_{a}<\eta_{b}$.
Thus, it does not come as a surprise that $\eta_{\perp}(\mathrm{ND})$ and
$\eta_{\parallel}(\mathrm{NC})$ take close values. This argument implies
moreover that diffusion in the ND phase should be faster perpendicular to the
director, i.e. $D_{\parallel}<D_{\perp}$ \cite{noteND}.

In summary, we have investigated diffusion in lyotropic LCs with two different
anchoring conditions. Our measurements confirm Stokes friction $D\propto1/d$
in both cases, unlike a previous study on thermotropic NLCs \cite{Ska10}. In
the NC phase, we found an unusually large viscosity ratio $\eta_{\bot}%
/\eta_{\Vert}\approx4$ which can be accounted for thanks to our perturbative
theoretical approach. Our analysis imposes $\eta_{b}<\eta_{a}<\eta_{c}$ on the
Miesovicz parameters in the NC phase (which is usually the case \cite{Sim03}),
and suggests $\eta_{c}<\eta_{a}<\eta_{b}$ with $D_{\parallel}<D_{\perp}$ in
the ND phase. As a short-term follow-up work, we will evaluate Eq. (\ref{16})
for the case of normal anchoring and an independent measurement of the
Miesovicz viscosities together with the additionnal parameter $\eta_{d}$ would
be most desirable.

We acknowledge financial support from the French National Research Agency
under grant \# ANR-07-JCJC-0023 and the Conseil R\'{e}gional d'Aquitaine.

\end{document}